\documentclass{article}
\usepackage[a4paper]{geometry}
\usepackage[english]{babel}

\usepackage[utf8]{inputenc}
\usepackage{xcolor}
\usepackage{authblk}

\usepackage{amsmath,amssymb}
\usepackage{hyperref}
\usepackage{stmaryrd}
\usepackage{todonotes}
\usepackage{cleveref}
\usepackage{multirow}

\usepackage{enumitem}

\newcommand{\tmop}[1]{\ensuremath{\operatorname{#1}}}

\newtheorem{remark}{Remark}
\newtheorem{proposition}{Proposition}
\newtheorem{lemma}{Lemma}
\newtheorem{corollary}{Corollary}
\newtheorem{definition}{Definition}
\newtheorem{theorem}{Theorem}

\newenvironment{proof}{\paragraph{Proof:}}{\hfill$\square$}

\usepackage{algorithm}
\usepackage{algpseudocode}

\definecolor{mybestcolor}{HTML}{BF3131}

\title{Reading Rational Univariate Representations on lexicographic Gr{\"o}bner bases.}
    \author[1]{Alexander Demin\thanks{\url{asdemin_2@edu.hse.ru}}}
    \author[2]{Fabrice Rouillier\thanks{\url{Fabrice.Rouillier@inria.fr}}}
    \author[2]{Jo\~{a}o Ruiz\thanks{\url{joao.ruiz@imj-prg.fr}}}
    \affil[1]{National Research University, Higher School of Economics, Moscow, Russia}
    \affil[2]{Sorbonne Université, Paris Université, CNRS (Institut de Mathématiques de Jussieu Paris-Rive-Gauche), Inria, Paris, France}

\newcommand\KK{\mathbb{K}}

\newcommand\deff{:=}

\newcommand\B{\mathcal{B}}

\begin{document}
    \maketitle
    
    \section{Introduction}
    Let $\KK$ be an effective field and $I\subseteq\KK[X_1,\ldots,X_n]$ a zero-dimensional ideal. 
    Let $\overline{\KK}$ be a fixed algebraic closure of $\KK$ and consider the set of zeros of $I$ in $\overline{\KK}$ defined as
    $
        V(I) = \{x \in \overline{\KK}^n, f(x) = 0 ~\text{for all}~f \in I\}.
    $
    The aim of this paper is finding parametrizations of $V(I)$ in the form
    \begin{equation*}
        \label{eq:param}
        f(T) = 0, X_1 = \frac{g_1(T)}{g(T)}, \ldots, X_n = \frac{g_n(T)}{g(T)},
    \end{equation*}
    where $f, g, g_1,\ldots,g_n \in \KK[T]$.
    
    We propose a Las Vegas algorithm for calculating such a parametrization for arbitrary zero-dimensional ideals using Gr\"obner bases. 
    By Las Vegas, we mean an algorithm that always produces a correct result but with a probabilistic complexity.

    Such parametrizations can be defined through Chow forms. Let $R(U_0,U_1,\ldots,U_n)=\prod_{\alpha=(\alpha_1,\ldots,\alpha_n)\in V(I)}(U_0-(\alpha_1U_1+\cdots+\alpha_nU_n))^{\mu(\alpha)}\in\KK[U_0,\ldots,U_n]$ where for $\alpha\in V(I)$, $\mu(\alpha)$ is its multiplicity, let $\overline{R}$ be the square-free part of $R$ and set $R_i\deff\frac{\partial}{\partial U_i}R$, for $i=0,1,\ldots,n$. If we choose $(t_1,\ldots,t_n)\in\KK^n$ such that $\overline{R}(U_0,t_1,\ldots,t_n)$ is squarefree, or, equivalently, such that the linear form $\sum_{i=1}^nt_iX_i$ is injective on $V(I)$, then $R(U_0,t_1,\ldots,t_n)=0$ and $X_i=\frac{R_i(U_0,t_1,\ldots,t_n)}{\frac{\partial}{\partial U_0}R_0(U_0,t_1,\ldots,t_n)}$ for $i = 1,\ldots,n$ is a parametrization of the zeroes of $I$.
    
    \begin{definition}[Rational Univariate Representation]
    \label{defRUR}
    Given $(t_1,\ldots,t_n)$ from $\mathbb{K}^n$, the polynomials $R(U_0,t_1,\ldots,t_n)$, $R_i(U_0,t_1,\ldots,t_n),i=1\ldots n\in \KK[X_1,\ldots,X_n]$ are uniquely defined. These polynomials define the Rational Univariate Representation (RUR) of $V(I)$ associated to $t$ when $t=\sum_{i=1}^nt_iX_i$ is injective on $V(I)$ (we then say that $t$ separates $V(I)$), and a RUR-Candidate for $V(I)$ otherwise.
    \end{definition}
    
    Starting with $I$, the computation of a parametrization in this way thus consists in solving {\em both} of the following problems:

    

    \begin{itemize}
    \item {\bf Problem S}: finding a separating linear form $t=\sum\limits_{i=1}^{n} t_iX_i$, that is, an $n$-tuple $(t_1,\ldots,t_n)\in\KK^n$ that does not cancel the discriminant of $\overline{R}$ with respect to $U_0$;
    \item {\bf Problem C}: given any 
    $n$-tuple $(t_1,\ldots,t_n)$ from $\KK^n$, 
    computing the polynomials $f_T \deff R(U_0,t_1,\ldots,t_n)$ and $f_{T,i} \deff R_i(U_0,t_1,\ldots,t_n)$ for $i=1,\ldots,n$, which define the candidate parametrization. 
    \end{itemize}

   The polynomial $R(U_0,U_1,\ldots ,U_n)$ can be computed using multivariate resultants in the case of systems without points at infinity \cite{Re92}. Roughly speaking, given a system of equations in $\mathbb{K}[X_1,\ldots,X_n]$ and $n+1$ independent variables $U_0,U_1,\ldots , U_n$, this $R\in \mathbb{K}[U_0,\ldots ,U_n]$ can be obtained by factorizing the $U$-resultant of the system. In the general case where there might be a positive-dimensional component at infinity, \cite{Can87} proposes a notion of Generalized Characteristic Polynomial, adding a new variable $s$ and some new polynomials to obtain a polynomial in $\mathbb{K}[U_0,\ldots ,U_n][s]$ and extract $R(U_0,U_1,\ldots ,U_n)$ from its coefficient of lower degree in $s$.
   
    These algorithms are difficult to program efficiently, but they ensure that the polynomials $f_T,f_{T,1},\ldots,f_{T,n}$ are computable and have well controlled sizes, since $R$ is a factor of the determinant of a certain matrix whose entries are the coefficients of the polynomials defining the studied ideal (the so-called Macaulay Matrix). For example, in \cite{EMT10} the authors show that if the initial system has polynomials with of degree at most $d$ with integer coefficients of bitsize at most $\tau$, then there exist some separating linear forms with integer coefficients such that the polynomials $f_T,f_{T,1},\ldots,f_{T,n}$ have coefficients of bitsize $\tilde{O}(nd^{n-1}\tau)$, where we use the shorthand $\tilde{O}(g(n)) = O(g(n) (\log g(n))^{O(1)})$.
    
    Most implemented algorithms for computing parametrizations are based on the knowledge of the quotient $\frac{\mathbb{K}[X_1,\ldots X_n]}{I}$, {\it i.e.}, a monomial basis ${\cal B}=\{w_1,\ldots,w_D\}$ and the multiplication matrices $M_{X_1},\ldots ,M_{X_n}$ by each of the variables in the quotient. These data can be computed from a Gröbner basis in $O(nD^3)$ operations in $\KK$ in general and will be the framework of the present article (and in $O(nD^\omega)$ for some subclasses of ideals, where $\omega$ is such that multiplying two $m$ by $m$ matrices over $\KK$ takes $O(m^\omega)$ -- see \cite{Neig2016}).

    \subsection{State of the art}
    
    In \cite{ABRW96}, the authors use a monomial basis $\B$ and the multiplication matrices $M_{X_1},\ldots,M_{X_n}$
    to compute $R(U_0,t_1,\ldots,t_n)$ as well as the multiplicities $\mu(\alpha)$ for $\alpha\in V(I)$ directly from an arbitrary choice of $(t_1,\ldots,t_n) \in \KK^n$, and propose computable parametrizations for each multiplicity. 
    
    In \cite{Rouillier:inria-00098872}, the author consolidates the strategy from \cite{ABRW96}, showing that all the polynomials $R(U_0,t_1,\ldots,t_n)$ and $R_i(U_0,t_1,\ldots,t_n)$ for $i=1,\ldots,n$ can be deduced from the traces of the matrix of multiplication by $X_jT^k$ in $\frac{\KK[X_1,\ldots,X_n]}{I}$ with respect to $\B$, that is, $\operatorname{Tr}\left(M_{X_jT^k}\right)$, with $j=1,\ldots,n$ and $k=1,\ldots,D$, and then proposing a Las Vegas algorithm that solves both problems {\bf S} and {\bf C}. This algorithm chooses a form $T=\sum_{i=1}^nt_iX_i$, computes ${\cal T}=\{w_iw_j,w_i,w_j\in {\cal B}\}$ in $O(\sharp{\cal T}D^2)$ operations in $\mathbb{K}$, then computes a RUR-Candidate in $O(D^3+nD^2)$ operations in $\mathbb{K}$. It finishes by checking that $T$ is separating within $O(D^3+nD^2)$ arithmetic operations in $\mathbb{K}$. While the computation of the RUR-Candidate can be improved to $O(nD^{5/2})$ by using \cite{Bostan} with an {\it a posteriori} check with an overhead $O(n\sharp{\cal T}D^{3/2})$ (see \cite{rouillier:tel-01435142}), the bottleneck is still the computation of ${\cal T}$ in $O(\sharp{\cal T}D^{2})$ with $\sharp{\cal T}\leq \operatorname{min}(2^nD,D^2)$. 
    
    In the case of ideals in shape position, {\it i.e.}, ideals for which the minimal polynomial of $M_{X_n}$ is equal to its characteristic polynomial, a simple modification of the FGLM algorithm allows to obtain rational parametrizations with a deterministic and efficient execution in $\tilde{O}(D^3+nD^2)$ (it solves problems {\bf S} and {\bf C} or returns a message of failure). In \cite{FM17}, this complexity drops even further to $\tilde{O}(D(N_1+nD))$ where $N_1$ is the number of non-zero elements of $M_{X_1}$, always with a deterministic algorithm and the complexity drops again to $\tilde{O}(D(N_1+n))$ with a probabilistic algorithm.
    
    When the ideal is not in shape position, one can swap the variables or perform a linear change of variables with the hope of falling back to an ideal in shape position. This is possible if and only if $\frac{\mathbb{K}[X_1,\ldots X_n]}{I}$ is cyclic (which is still not the general case). Testing if $\frac{\mathbb{K}[X_1,\ldots X_n]}{I}$ is cyclic is equivalent to finding a primitive element $T$ in $\frac{\mathbb{K}[X_1,\ldots X_n]}{\sqrt{I}}$, for example a linear form $T=\sum_{i=1}^nt_iX_i$ that is injective on $V(I)$ and then checking if the minimal polynomial of $M_T$ in $\frac{\mathbb{K}[X_1,\ldots X_n]}{I}$ has degree $D$. 

    It is easy to adapt FGLM variants into a Las Vegas algorithm for computing parametrizations in the case of ideals with cyclic quotient algebras, in the same complexity as for the shape position case. If the ideal is known to be cyclic in advance, solving problem {\bf S} consists in trying linear forms until finding a linear form whose minimal polynomial has degree $D$. If it is not known to be cyclic in advance, problem {\bf S} cannot be solved in this way, and the algorithm might return a wrong result.
    
    When the quotient algebra is not cyclic, a straightforward solution for solving problems {\bf S} and {\bf C} consists in first computing the radical of the ideal with a deterministic algorithm in order to fall back to the cyclic case with the above Las Vegas algorithm, but its complexity would then be prohibitive compared to the solution proposed in \cite{Rouillier:inria-00098872}. Instead, in \cite{FM17}, a Monte Carlo variant sacrificing the resolution of problem {\bf S} is proposed with a complexity in $\tilde{O}(\sqrt{n}D^{3-\frac{1}{n}})$. The authors of \cite{Neiger} improved this Monte Carlo algorithm exploiting some duality concepts from  \cite{Bostan} to get a complexity in $O(D^3+nD^2)$ for problem {\bf C}.
   
   Outside our framework (multiplication matrices as input) there exist very few alternatives. Among them, the algorithm from \cite{GLS01} is a Monte-Carlo alternative that solves problem {\bf C} starting directly from the initial system with a good efficiency in practice.

    \subsection{Our contributions}

    Our main contribution is a new Las Vegas algorithm for solving both problems {\bf S} and {\bf C} for any kind of zero-dimensional ideal. Let $t$ be a separating linear form, $D$ be the number of zeroes of the system counted with multiplicities (or, equivalently, the degree of the characteristic polynomial of $t$), and $\delta$ be the degree of the minimal polynomial of $t$.
    Our algorithm solves problems {\bf S} and {\bf C} in $\tilde{O}(D^2\delta+nD^2(D-\delta+1))$ arithmetic operations from the knowledge of a monomial basis ${\cal B}=\{w_1,\ldots,w_D\}$ of $\frac{\mathbb{K}[X_1,\ldots X_n]}{I}$ and the multiplication matrices $M_{X_1},\ldots ,M_{X_n}$ by each variable with respect to ${\cal B}$. In a parallel version one can exploit the fact that all the $f_{T,i},\,i=1,\ldots,n$ can be computed independently. 
    
    In the case of ideals with cyclic quotient, $D=\delta$ and the complexity drops to $\tilde{O}(D^2\delta+nD^2)$ and, moreover, this bound improves the best known complexity for solving problems {\bf S} and {\bf C} in the general case.
    
    However, when the input is a Gröbner basis, this computation shares the same upper bound in the worst case as the computation of its prerequisite (the multiplication matrices which are computed in $O(nD^3)$ operations in $\mathbb{K}$) when the input is a Gröbner basis and thus the same complexity as the state of the art Monte-Carlo algorithms that solve only problem {\bf C} using Gröbner bases.
    
    Whenever convenient, in the proofs we will assume that the characteristic of $\KK$ is either zero or sufficiently large.
    
    In the experimental part, we measure the influence of the choice of the separating form on the computations (sparsity of the matrices, size of the result) and propose some strategies that tend to produce {\em good} candidates.


    
    \section{Lexicographic Gr\"obner bases and RUR candidates}
\label{section:2}

In this section we will focus on solving problem {\bf C}, leaving problem {\bf S} to the next section. 

Let $I\subseteq\KK[X_1,\ldots,X_n]$ be a zero-dimensional ideal and $t=\sum_{i=1}^n t_iX_i$ a linear form with coefficients in $\KK$. 

The goal of this section is to compute a set of polynomials $h(T),h_{1}(T),\ldots ,h_n(T)$ such that   $h(T)=0$ and $X_i=\frac{h_i(T)}{h'(T)}$, $i=1,\ldots,n$, define the RUR-Candidate of $V(I)$ associated to a given linear form $t$. 

Put $I_T\deff I+\left\langle T-t\right\rangle\subseteq\KK[T,X_1,\ldots,X_n]$, still a zero-dimensional ideal but now with one more variable, called here the augmented ideal. We note that $\frac{\KK[T,X_1,\ldots,X_n]}{I_T}$ and $\frac{\KK[X_1,\ldots,X_n]}{I}$ are isomorphic as vector spaces and can share the same monomial basis.

\begin{definition}
\label{abusenotation}
The polynomials $h(T),h_1(T),\ldots, h_n(T)$ are called a RUR (resp. RUR candidate) of $V(I)$ associated to the separating linear form (resp. linear form) $t$, and we will also call them a RUR (resp. RUR candidate) of $V(I_T)$ associated to the variable $T$, with the missing coordinate function in the second case being $\frac{Th'(T)}{h'(T)}$. Also, a RUR (RUR candidate) associated to a variable is an $n$-uple, while a RUR (RUR candidate) associated to a linear form is an $n+1$-uple.
\end{definition}

In a RUR $h(T),h_{1}(T),\ldots ,h_n(T)$ associated to a linear form $t$, $h(T)$ is the (monic) characteristic polynomial of the multiplication by $t$ in $\frac{\mathbb{K}[X_1,\ldots,X_n]}{I}$. Taking $g(T)=\gcd(h(T),h'(T))$ one gets that $\frac{\tilde{h}_i(T)}{\tilde{h'(T)}}=\frac{h_i(T)}{h'(T)}$ with $\tilde{h'}=\frac{h'}{g}$ and $\tilde{h}_i=\frac{h_i}{g},i=1,\ldots,n$.

\begin{definition}
\label{reducedRUR}
With the above notations, we set $f_i(T)=\frac{1}{\deg(h)}\tilde{h}_i$ and $f_0(T)=\frac{1}{\deg(h)}\tilde{h'}$. In particular $f_0$ is monic, proportional to the squarefree part of $h'$ and we have $\frac{f_i}{f_0}=\frac{h_i}{h'}$.
Then, we define $h,f_1,\ldots f_n$ as the reduced RUR (resp. reduced RUR-Candidate) of $V(I)$ associated to the same variable or linear form as the RUR (response.  RUR-Candidate) $h,h_1,\ldots, h_n$.
\end{definition}

\begin{remark}
\label{RURfromradical}
The idea for introducing the reduced RUR is that the passage from a reduced RUR of $V(I)$ to one of $V(\sqrt{I})$ both associated to the same linear form is easy: doing so consists of replacing the first polynomial $f$ with $h$, and this polynomial can be computed on $\tilde{O}(D^3)$ arithmetic operations.
\end{remark}

Given a linear form $t$, our algorithm will first {\em read} ({\it i.e.} compute with a closed formula) a parameterization of the solutions on some lexicographic Gröbner basis of $I_T$, say polynomials $\tilde{f}(T),h_{i,1}(T),h_{i,0}(T),i=1\ldots n$ such that $\tilde{f}(T),h_{i,1}(T)X_i+h_{i,0}(T),i=1\ldots n$ is a parameterization of $V(I_T)$ associated to the variable $T$ (or equivalently a parameterization of $V(I)$ associated to the linear form $t$) whenever $T$ (or equivalently $t$) separates the system.

Obtaining a zero-dimensional parametrization relates directly with obtaining a RUR since we can easily pass from one to the other:

    \begin{corollary}
    \label{paramtoRUR}
        Let $\tilde{f}(T),h_{i,1}(T)X_i+h_{i,0}(T),i=1\ldots n$ be a parameterization of $V(I)$ associated to some separating linear form $t$ with $\tilde{f}$ monic. Set $\overline{f}(T)$ the squarefree part of $\tilde{f}(T)$ and $f_{X_i}=\frac{-1}{\deg(\overline{f})}h_{j,0}(T)h_{j,1}(T)^{-1}\overline{f}'\mod\overline{f}$. 
        Then $(\overline{f},f_{X_1},\ldots,f_{X_n})$ is the reduced RUR of $V(\sqrt{I})$ associated to $t$.
        Moreover, if $f$ is the characteristic polynomial of the multiplication by $t$ in  $\frac{\mathbb{K}[X_1,\ldots,X_n]}{I}$, then $(f,f_{X_1},\ldots,f_{X_n})$ is the reduced RUR of $V(I)$ associated to $t$.
    \end{corollary}

 \subsection{Reading parameterizations candidates on lexicographic Gröbner basis}
 
    Given a zero dimensional ideal $I$, our global strategy for the computation of a RUR-Candidate of $V(I)$ associated to a given linear for $t$ consists in computing the RUR-Candidates associated to the variable $T$ of the $n$ subsets $V((I_T)\cap \KK[T,X_i])$ to get separately the coordinate functions.
    
    
 \subsubsection{The bivariate case}
 
  In the bivariate case, the strategy is almost word for word lifted from \cite{BLPR15} and \cite{lazard:hal-01468796}, which uses a subresultant sequence to compute a rational parametrization when one of the two variables separates the solutions. Here, instead of subresultants, we use a lexicographic Gr\"obner basis. Note, however, that in the case of systems defined by two polynomials, the polynomials appearing in all calculations end up being {\it the same}.

    \begin{remark}\label{remarksubresultants}
        One may wonder if the parallel between subresultant sequences and Gr\"obner bases is a sign of a more general phenomenon. Unfortunately we couldn't find relations beyond what has been already found, relating ideals in shape position to multivariate subresultants in \cite{Cox23}. Even in the bivariate case, not necessarily the subresultant polynomials are equal to the polynomials in a lexicographic Gr\"obner basis. For example, given two polynomials $f,g\in\KK[X,Y]$, a (reduced) Gr\"obner basis of $\left\langle f,g\right\rangle$ with $X>Y$ is of the form $\{p(Y),g_1(X,Y),\ldots,g_n(X,Y)\}$. However, the resultant of these two polynomials with respect to $X$, noted $\operatorname{Res}_X(f,g)(Y)=\operatorname{SRes}_{X,0}(f,g)(Y)$ only needs to be {\it a multiple} of $p(Y)$, not necessarily equal to it (nor equal up to a scalar multiple).
    \end{remark}
    
    \paragraph{}The link with the subresultants based algorithm from \cite{BLPR15} comes from \cite{Gia89}:
    \begin{proposition}[Last corollary in \cite{Gia89}]\label{propositionresultgianni}
        Let $I\subseteq\KK[x_1,\ldots,x_n]$ be a zero-dimensional ideal, $G=(g_1,\ldots,g_s)$ be its Gr\"obner basis w.r.t. purely lexicographic ordering with $x_1>\cdots >x_n$ and let $(\alpha_k,\ldots,\alpha_n)\in\overline{\KK}^{n-k+1}$ be a zero of $I\cap\KK[x_k,\ldots,x_n]$. Let $I_k=(G\cap\KK[x_{k-1},\ldots,x_n])\backslash\KK[x_k,\ldots,x_n]=(g_{k_1},\ldots,g_{k_r})$. Let $g\in I_k$ be a polynomial of minimal degree in $x_{k-1}$ whose leading coefficient $c(x_k,\ldots,x_n)$ doesn't vanish when evaluated in $(\alpha_k,\ldots,\alpha_n)$. Then:
        \begin{equation*}
            g_\alpha\deff g(x_{k-1},\alpha_k,\ldots,\alpha_n) = \gcd (g_{k_1}(x_{k-1},\alpha_k,\ldots,\alpha_n),\ldots,g_{kr}(x_{k-1},\alpha_k,\ldots,\alpha_n)). 
        \end{equation*}

        In particular, if $\alpha_{k-1}$ is a zero of $g_\alpha$, then $(\alpha_{k-1},\ldots,\alpha_n)$ is a zero of $I\cap\KK[x_{k-1},\ldots,x_n]$. 
    \end{proposition}
    
    What this result tells us is that, under a specific phrasing, Gr\"obner bases {\it do} specialize similarly to subresultants, which is why we can generalize a subresultant based strategy to Gr\"obner bases. 
    
    \begin{proposition}\label{propositionmaintheorembivariate}
        Let $J\subseteq\KK[T,X]$ be a zero-dimensional ideal and $G$ a reduced lexicographic Gr\"obner basis of $I$ with $T<X$. 
        
        \noindent Without loss of generality, write $G=\{f(T),g_{k}(T,X)|\,k=1,\ldots,m\}$ where $g_k(T,X) = \sum_{i=1}^k a_{k,i}(T)X^i$ is either $0$ or has degree exactly $k$ ($a_{k,k}\neq 0$). 
        
        \noindent Let $h_0 = \overline{f}$ and for each $k=1,\ldots,m$, let $h_k=\gcd(h_{k-1},a_{k,k})$ and $f_k = \frac{h_{k-1}}{h_{k}}$. 
        
        If $T$ separates $V(J)$, then
            \begin{equation*}
                V(J) = \cup_k V(f_k(T),ka_{k,k}(T)X+a_{k,k-1}(T)).
            \end{equation*}
    \end{proposition}
    \begin{proof}
        Given a root $\alpha$ of $f$, let $n_\alpha$ be the first index $k$ such that $a_{k,k}(\alpha)\neq0$. Note that for each $k$, the $f_k$ defined in the statement consists of the squarefree polynomial whose roots are the $\alpha\in V(f)$ for which $n_\alpha=k$: since the roots of $h_k$ are exactly the roots of $f$ that kill $a_{1,1},\ldots,a_{k,k}$, then $f_k$ has roots that kill $a_{1,1},\ldots,a_{k-1,k-1}$ but {\bf not} $a_{k,k}$. Because of this, if $\alpha$ is a root of $f_k$,
        \begin{equation*}
            g_{k}(\alpha,X) = \gcd\left(g_1(\alpha,X),\ldots,g_m(\alpha,X))\right)
        \end{equation*}

        from proposition \ref{propositionresultgianni}. This means that the roots of $g_{n_\alpha}(\alpha,X)$ are $X$ coordinates of roots of the system whose $T$ coordinate is $\alpha$, and reciprocally all roots of the system whose $T$ coordinate is $\alpha$ have an $X$ coordinate that kills $g_{n_\alpha}(\alpha,X)$. 

        Since we're supposing that $T$ separates the system, there is exactly one solution with $T$ coordinate equal to $\alpha$. In other words, $g_{n_\alpha}(\alpha,X)$ has exactly one root and should be of the form $b_{n_\alpha}(X-\beta)^{n_\alpha}$. So for all $\alpha$ such that $n_\alpha=k$ for a fixed $k$ (in other words, let $\alpha$ be a root of $f_k$),
        \begin{equation*}
            g_k(\alpha,X) = a_{k,k}(\alpha)(X-\beta)^k,
        \end{equation*}

        and computing the parametrization of $X$ is equivalent to computing an expression for this $\beta$. However we know that $g_k(T,X)=\sum_i a_{k,i}(T)X^i$, so that the coefficient of $X^{k-1}$ has to be
        \begin{equation*}
            a_{k,k-1}(\alpha) = -ka_{k,k}(\alpha)\beta,
        \end{equation*}

        which finally gives $\beta$ as the only root of $ka_{k,k}(\alpha)X+a_{k,k-1}(\alpha)=0$. Since this equality is true for all roots $\alpha$ of $f_k$, thanks to the Chinese Remainder Theorem (and because $f_k$ is squarefree), we have that
        \begin{equation*}
            V(f_k(T),ka_{k,k}(T)X+a_{k,k-1}(T)) \subseteq V(J). 
        \end{equation*}

        Moreover, since every $\alpha$ root of $f$ has a well-defined $n_\alpha$ (because the system is zero-dimensional, so $a_{m,m}$ has to be a nonzero scalar and therefore any $\alpha$ that killed all leading coefficients up to that point cannot kill $a_{m,m}$), we also have that any element of $V(J)$ is also in {\it some} $V(f_k(T),ka_{k,k}(T)+a_{k,k-1}(T))$, which establishes that
        \begin{equation*}
            V(J) = \cup_k V(f_k(T),ka_{k,k}(T)X+a_{k,k-1}(T)). 
        \end{equation*}
    \end{proof}
    
    \begin{remark}\label{remarkdecompositionbivariate}
        This method decomposes the system according to the multiplicity of the coordinate $X$: it induces a somewhat natural decomposition of the system in two variables, that is organized according to the total multiplicity of the $X$ coordinate of the solutions. That is, each factor $f_k$ contains all solutions whose $X$ coordinate has multiplicity equal to $k$. 
    \end{remark}

    \paragraph{}We can finish solving problem {\bf C} by putting all these partial parametrizations together into a single one of the entire system:

    \begin{corollary}
        Using the same notations as in proposition \ref{propositionmaintheorembivariate}, we define the parametrization candidate $f(T),\,h_{X,1}(T)X+h_{X,0}(T)$ of $V(J)$ with
        \begin{align*}
            h_{X,0}(T) \deff \sum_k a_{k,k-1}(T)\prod\limits_{i=1}^{k-1}f_i(T)\mod h_0\\
            h_{X,1}(T) \deff \sum_k ka_{k,k}(T)\prod\limits_{i=1}^{k-1}f_i(T)\mod h_0
        \end{align*}
        so that, if $T$ separates $J$, we have the following parametrization associated to the variable $T$: $\{f(T)=0,\,h_{X,1}(T)X+h_{X,0}(T)=0\} = V(J)$.
    \end{corollary}

    \paragraph{}This result gives rise to Algorithm~\ref{algorithmrationalparametrizationbivariate} in two variables. 
    \begin{algorithm}\caption{BivariateParametrizationCandidate}\label{algorithmrationalparametrizationbivariate}
        \begin{algorithmic}[1]
            \Require A lexicographic Gr\"obner basis
            \State $G=\{f(T),g_1(T,X),\ldots,g_{m}(T,X)\}$ of $J\subseteq\KK[T,X]$ with
            \State $g_k(T,X)=\sum\limits_{i=0}^{\partial_k}a_{k,i}(T)X^i$ (possibly with some $g_k=0$). 
            \Ensure A parametrization-candidate of $V(J)$ associated to the variable $T$, $\{\overline{f},h_{X,1}X+h_{X_0}\}$.
            \State Compute $h_0\deff\overline{f}$ the squarefree part of $f$
            \State $\rho\deff1$, $h_{X,0}\deff0$, $h_{X,1}\deff0$
            \For{$k=1,\ldots,m$}
                \State $h_k\deff\gcd(h_{k-1},a_{k,\partial_k})$, $f_k\deff\frac{h_{k-1}}{h_k}$
                \State $\rho\deff\rho\cdot f_k$
                \State $h_{X,1}\deff h_{X,1}+\partial_ka_{k,\partial_k}\rho\mod h_0$
                \State $h_{X,0}\deff h_{X,0}+a_{k,\partial_k-1}\rho\mod h_0$
            \EndFor
        \end{algorithmic}
    \end{algorithm}

    \subsection{The computation of the required bivariate lexicographic Gröbner bases}

 Let's now focus on the computation of the bivariate Gr\"{o}bner bases required for Algorithm \ref{algorithmrationalparametrizationbivariate} in our context, that is to say the computation of lexicographic Gröbner basis of $I\cap\mathbb{K}[T,X_i],i=1\ldots n$ from the knowledge of a monomial basis ${\cal B}$ of $\frac{\mathbb{K}[T,X_1,\ldots ,X_n]}{I}$. For that purpose, we use the first steps of {\em FGLM} algorithm computing separatly the first univariate polynomial (Algorithm \ref{minimalpolynomialalgo}) since it will be shared by all the bivariate bases we need and the bivariate polynomials (Algorithm \ref{coordinatealgo}).

    \begin{algorithm}\caption{MinimalPolynomial}\label{minimalpolynomialalgo}
        \begin{algorithmic}[1]
            \Require $J\subseteq \mathbb{K} [T, X_1,..,X_n]$ a zero-dimensional ideal
            \State $M_T$ the matrix of multiplication by $T$ in $\frac{\mathbb{K} [T, X_1,..,X_n]}{J}$ in any monomial basis $\mathcal{B}_0$
            \Ensure The minimal polynomial $f$ of $M_T$ and 
            \begin{itemize}
                \item ${\cal B}=\{1,T,\ldots, T^{\delta-1} \}$ where $\delta = \tmop{deg} (f)$
                \item The vectors $\overrightarrow{1}, \overrightarrow{T}, \ldots, \overrightarrow{T^{\delta - 1}}$ in $\mathcal{B}_0$
                \item The upper triangular matrix $U$ in row echelon form of the vectors $1,T, \ldots, T^{\delta - 1}$
                \item The lower triangular matrix $L$ of the change of basis for getting the row echelon form of $U$
            \end{itemize}
            \State $D\deff \sharp\B_0$; $U := 0_D$; $L:=\tmop{Id}_D$.
            \State Push $[1, 0, \ldots, 0]$ in $U$ (in replacement of the first line)
            \State $i:=0$; $\delta:=0$;
            \While{$\delta = 0$}
                \State Compute $\overrightarrow{T^{i + 1}} \deff M_T  \overrightarrow{T^i}$
                \State Push $\overrightarrow{T^{i + 1}}$ in $U$ (at the place of the first null line - denoted $\operatorname{index}(T^{i + 1})$), reduce $U$ (to upper triangular form) and update $L$ (same operations on $L$ as done on $U$)
                \State If $\mathcal{B} \cup \{ T^{i + 1} \}$ is free, then $\tmop{Append} (T^{i+1}, \mathcal{B})$
                \State else $\tmop{Append} \left( T^{i + 1} - \sum^{\sharp \mathcal{B}}_{j=1} (L[\operatorname{index}(T^{i+1})][j])T^j, G \right)$, and set $\delta := i$
                \State $i:=i + 1$
            \EndWhile
        \end{algorithmic}
    \end{algorithm}

    \begin{algorithm}\caption{Coordinate}\label{coordinatealgo}
        \begin{algorithmic}[1]
            \Require $J\subseteq \mathbb{K} [T, X_1,..,X_n]$ a zero-dimensional Ideal
            \State $M_X$ the matrix of multiplication by $X$ in $\frac{\mathbb{K} [T,X_1,\ldots ,X_n]}{J}$ in any base $\mathcal{B}_0$ and the output of Algorithm \ref{minimalpolynomialalgo} call with the same ideal $J$ and the same base $\mathcal{B}_0$
            \Ensure A Gröbner basis $G$ of $I\cap \mathbb{K}[T,X]$.
            \State $k:=0$;~$\delta:=\sharp {\cal B}$;
            \While{$\delta > 0$}
                \State $k:=k+1$;
                \For{$j=0\ldots \delta$} 
                \State Compute $\overrightarrow{X^k T^j} := M_X  \overrightarrow{X^{k - 1} T^j}$
                \State Push $\overrightarrow{X^k T^j}$ in $U$, reduce $U$ and update $L$
                \State If $\mathcal{B} \cup \{ X^k T^j \}$ is free, then $\tmop{Append} (X^kT^j, \mathcal{B})$
                \State else $\tmop{Append} \left( X^kT^j - \sum^{\sharp \mathcal{B}}_{j=1} (L[\operatorname{index}(X^kT^j)][j]) \mathcal{B} [j], G \right)$, set $\delta := j$
                \EndFor
            \EndWhile
        \end{algorithmic}
    \end{algorithm}

\subsection{The computation of RUR-Candidates}
  
   Given any zero-dimensional ideal $I\subseteq \mathbb{K} [X_1,\ldots,X_n]$, any linear form $t=\sum_{i=1}^nt_iX_i$ and any monomial basis ${\cal B}$ of $\frac{\mathbb{K} [X_1,\ldots,X_n]}{I}$, ${\cal B}$ is also a monomial basis of $\frac{\mathbb{K} [T,X_1,\ldots,X_n]}{I+\langle T-t \rangle}$. In particular, the matrices of multiplications $M_{X_i}$ in these two algebra are all the same in this shared basis. Moreover, $\sum_{i=1}^nt_iM_{X_i}$ is the matrix of multiplication by $t$ in $\frac{\mathbb{K} [X_1,\ldots,X_n]}{I}$ and also the matrix of multiplication by the variable $T$ in $\frac{\mathbb{K} [T,X_1,\ldots,X_n]}{I+\langle T-t \rangle}$.
 
    Putting all together, one obtains Algorithm \ref{RadicalRURCandidatealgo} for computing RUR Candidates for $V(\sqrt{I})$ (which also allows to compute RUR-Candidates of $V(I)$ modulo the computation of a characteristic polynomial (see Remark \ref{RURfromradical}).
   
    \begin{algorithm}\caption{RadicalRURCandidate}\label{RadicalRURCandidatealgo}
        \begin{algorithmic}[1]
            \Require $M_{X_1},\ldots M_{X_n}$ the matrices of multiplication by each variable in $\frac{\mathbb{K} [X_1,\ldots,X_n]}{I}$ in some base $\mathcal{B}$ and a linear form $t=\sum_{i=1}^nt_iX_i\in\mathbb{K} [X_1,\ldots,X_n]$.
            \Ensure $\overline{f}(T)=0,X_i=\frac{f_i(T)}{\overline{f}'(T)}$, the reduced RUR-Candidate of $V(\sqrt{I})$ associated to $t$
            \State Compute $\tilde{f},data:=\operatorname{MinimalPolynomial}(I+\langle T-t \rangle,\sum_{i=1}^nt_iM_{X_i})$
            \State Compute $\overline{f}(T)$, the squarefree part of $\tilde{f}(T)$.
            \For{$i=1\ldots n$} 
                \State $G_i:={f}\cup \tmop{Coordinate}(I+\langle T-t \rangle,M_{X_i},data)$
                \State $h_{i,1}X_i+h_{i,0}:=\tmop{BivariateParametrizationCandidate}(G_i)$    
                \State $f_{X_i}:=\frac{-1}{\deg(\overline{f})}h_{j,0}(T)h_{j,1}(T)^{-1}\overline{f}'\mod\overline{f}$
            \EndFor
        \end{algorithmic}
    \end{algorithm}

    \subsection{Complexity analysis}
    
    We first begin with the computation of the parametrization of each variable, supposing that the chosen linear form does separate $V(I)$. Recall the notation that $I\subseteq\KK[T,X]$ is a zero-dimensional ideal and $G$ a reduced lexicographic Gr\"obner basis of $I$, with $T<X$. Because of the following Remark~\ref{nbgb}, we can write $G=\{f(T),g_k(T,X),k=1,\ldots,m\}$ where $g_k(T,X)=\sum_{j=0}^k a_{k,j}(T)X^j$ and we allow some of the $g_k$ to eventually be zero. 
    
    An observation that will come in handy throughout this paper is that we can control very well the multidegrees of the polynomials appearing in a lexicographic Gr\"obner basis of bivariate zero-dimensional ideals.
    
    \begin{remark}
    \label{nbgb}
      Let $G=\{f,g_k,\,k=1,\ldots,m\}$ be a reduced lexicographic ($X>T$) Gr\"obner basis of a zero-dimensional system $\left\langle G\right\rangle\subseteq\KK[T,X]$: then it is well known that $f$ is a univariate polynomial in $T$ (see the discussion in \cite[5\S3]{CLO91}). The next polynomial $g_1$ will have leading monomial $T^aX^b$ with $a<\deg_T(f)$ and $b>0$ (because of the hypothesis that $G$ is reduced). The one after that will have $\deg_T g_2 < \deg_T g_1$ and $\deg_X g_2 > \deg_X g_1$ and so on. Therefore the most terms such a Gr\"obner basis can have is $\deg_T f+1$ and in that case the sequence of multidegrees is
    \begin{equation*}
        \left(\deg_T(g_k),\deg_X(g_k)\right) = (\deg_T (f)-k,k), ~~k=0,\ldots,\operatorname{deg}_T(f). 
    \end{equation*}
    
    Moreover, the number of elements in such a Gr{\"o}bner basis does not exceed $2(D-\delta)+1$, where $\delta = \operatorname{deg}_T(f)$ and $D=\dim_\KK\frac{\KK[T,X]}{I}$. This can be observed by noting that in the enumeration of monomials in the FGLM algorithm there cannot be two consecutive monomials that belong to the leading ideal and differ in exactly one coordinate. Overall, we have that the number of elements in such a Gr{\"o}bner basis is bounded by $O(\operatorname{min}(\delta, D-\delta))$.
    
    \end{remark}
    
    \begin{lemma}[Complexity of the bivariate parametrization candidate]
        With the notation above, Algorithm \ref{algorithmrationalparametrizationbivariate} computes the parametrization candidate of $X$ associated to the choice of linear form in $\tilde{O}(m\delta)$ arithmetic operations in $\KK$, where $\delta=\deg_T (f)$. Bounding $m$ by $(D-\delta+1)$ accordingly to Remark \ref{nbgb}, we get $\tilde{O}((D-\delta+1)\delta)$ arithmetic operations in $\KK$. 
    \end{lemma}
    \begin{proof}
        The algorithm first computes $\leq D-\delta$ GCD's of polynomials of degree less than $\delta$. Therefore we arrive at a total of $\tilde{O}(\delta(D-\delta+1))$ arithmetic operations in $\KK$. The $+1$ appearing is to account for the radical case where $D=\delta$. 
    \end{proof}
    
    Note that when the quotient algebra is cyclic, we have $D=\delta$ and $m = 1$, and the above complexity drops to $\tilde{O}(\delta)$.
 
    
    

    If we now consider the multiplication matrices of each variable as our input, we then need to consider the complexity of computing the lexicographic Gröbner bases of $I_T\cap\mathbb{K}[T,X_i],i=1,\ldots,n$.
    
    The complexity of Algorithm \ref{minimalpolynomialalgo} is in $O(D^2\delta)$ since it performs $\leq\delta$ loops with a matrix-vector product ($O(D^2)$ each) and a row echelon operation for reducing one single row in the matrix ($O(D^2)$ each).

    The complexity of Algorithm \ref{coordinatealgo} is in $O((D-\delta+1)D^2)$ since it realizes at most $D-\delta+1$ loops (see Remark \ref{nbgb}) with a matrix-vector product ($O(D^2)$ each) and a row echelon operation for reducing one single row in the matrix ($O(D^2)$ each).

Putting all together, we get the following theorem~:
\begin{theorem}
Let $I\subseteq \mathbb{K}[X_1,\ldots, X_n]$ a zero-dimensional ideal, $D$ the number of its zeroes counted with multiplicities, and $t=\sum_{i=1}^nt_iX_i\in \mathbb{K}[X_1,\ldots, X_n]$ any linear form. 

The complexity of Algorithm \ref{RadicalRURCandidatealgo} for computing the RUR-Candidate of $V(\sqrt{I})$ associated to $t$ is $\tilde{O}(D^2(\delta + n(D-\delta+1)))$ operations in $\mathbb{K}$. Using Remark~\ref{RURfromradical} the computation the RUR-Candidate of $V(I)$ associated to $t$ can be done in  $\tilde{O}(D^2(D + n(D-\delta+1)))$.

In particular, when the quotient algebra $\frac{\mathbb{K}[X_1,\ldots, X_n]}{I}$ is cyclic, the complexities drop to $\tilde{O}(D^2(\delta + n))$ and $\tilde{O}(D^2(D + n))$, respectively.
\end{theorem}

 \subsection{RUR-Candidates from a unique Lexicographic Gröbner basis}

    Instead of using several elimination bases, it is possible to use a single, full lexicographic Gr\"obner basis. The procedure for the first variable is the bivariate case described above. And to obtain the next variable, we run an inductive procedure: suppose that $G$ is a lexicographic Gr\"obner basis of $I_T\subseteq\KK[T,X_1,\ldots,X_n]$ with $T<X_1<\cdots<X_n$. We can write $G=\{f(T)\}\cup\bigcup_{j=1}^n G_j$, where $G_j$ are the polynomials in $G$ whose largest variable is $X_j$. All polynomials in $G_j$ can be written as
    \begin{equation*}
        g_{j,k}(T,X_1,\ldots,X_j) = \sum_{i=1}^{n_{j,k}} a_{j,k,i}(T,X_1,\ldots,X_{j-1})X_j^i.
    \end{equation*}

    Suppose that we have all parametrizations up to a certain variable, so $f(T)=0$ and $X_i=p_i(T)$ for $i=1,\ldots,j-1$. The parametrization of $X_j$ can be obtained by substituting all $p_i$ into the $g_{j,k}$, so computing the set 
    \begin{equation*}
        \{f(T)\} \cup \{g_{j,k}(T,p_1(T),\ldots,p_{j-1}(T),X_j)\} \subseteq \KK[T,X_j]. 
    \end{equation*}

    It is important to note that this set is {\bf not} a Gr\"obner basis of $I_T\cap\KK[T,X_j]$. However, if $T$ separates $V(I)$ then Algorithm~\ref{algorithmrationalparametrizationbivariate} still computes a parametrization of $X_j$, mostly because Proposition~\ref{propositionresultgianni} still applies. 
    
    The first interest for considering such a strategy would be to have less polynomials coming from Gröbner bases to consider during the computation. Unfortunately, this is not the case since the number of elements of a zero-dimensional lexicographic Gröbner basis is in $O(nD)$ and is of the same order than the number of polynomials in our $n$ bivariate lexicographic Gröbner bases.

    
    
    
    \section{Separating elements and RUR}
\label{section:3}

    In the introduction, we addressed problem {\bf S} of finding a separating linear form. It can be further broken down into two parts: first we need to {\it choose} a linear form, and then we need to {\it check} whether or not the separation hypothesis is verified. 
    
    For the first part one might use the set proposed in \cite{ABRW96} which contains at least one separating linear form~:
    \begin{equation*}
        {\cal S}\deff\left\{\sum_{i=1}^n j^iX_i,\,j=1,\ldots,(n-1)\frac{D(D-1)}{2}\right\}
    \end{equation*}
    
    For the second part, separating linear forms $t$ will be those for which the degree of the squarefree part of the minimal polynomial of $M_t$ in $\frac{\mathbb{K}[X_1,\ldots , X_n]}{I}$ is maximal (and equal to $\sharp V(I)$). So, computing the minimal polynomial of each $M_t, t\in {\cal S}$ and keeping one with a squarefree part of maximal degree is a solution, but it is obviously inefficient.
    
    Alternatively, if ${\cal B}=\{w_i,i=1\ldots D\}$ is a basis of $\frac{\mathbb{K}[X_1,\ldots , X_n]}{I}$, one could compute Hermite's generalized quadratic form  $Q_1=\left(\operatorname{Tr}(M_{w_iw_j}))\right)_{i,j=1..D}$ whose rank equals $\sharp V(I)$ (see \cite{Rouillier:inria-00098872}). Then the computation of all $(n-1)\frac{D(D-1)}{2}$ minimal polynomials can be avoided, but this would require to compute all products ${\cal T}=\{w_iw_j,w_i,w_j\in {\cal B}\}$ which is the bottleneck in the algorithms from \cite{Rouillier:inria-00098872} and \cite{Bostan} as recalled in the introduction (the complexity is in $O(\sharp {\cal T}D^2)$ with $\sharp {\cal T}\leq \min(2^nD,D^2)$).
    
    Since our algorithm for computing RUR-Candidates is based on bivariate RUR-Candidates, let's begin looking at the second part in the case of bivariate systems with a strategy inspired from \cite{BLPR15}:

    \begin{proposition}[Bivariate separation check]
        Let $I\subseteq\KK[T,X]$ be a zero dimensional system and $G$ a reduced lexicographic Gr\"obner basis of $I$ with $T<X$. Without loss of generality, $G=\{f(T),g_k(T,X),\,k=1,\ldots,m\}$ with $g_k = \sum_{i=0}^k a_{k,i}(T)X^i$ and we allow for some of the $g_k$ to be zero. Let $h_0 = \overline{f}$ and for each $k=1,\ldots,m$, let $h_k=\gcd(h_{k-1},a_{k,k})$ and $f_k = \frac{h_{k-1}}{h_{k}}$. Then $T$ separates $V(I)$ if and only if for every $k=1,\ldots,m$ and $i=0,\ldots,k-1$,
        \begin{equation*}
            \frac{k-i}{i+1}ka_{k,k}a_{k,i}\equiv a_{k,i+1}a_{k,k-1}\mod f_k
        \end{equation*}
    \end{proposition}
    \begin{proof}
        Recall from the proof of Proposition~\ref{propositionmaintheorembivariate} that if $\alpha$ is a root of $f_k$, the roots of $g_k(\alpha,X)$ are the $X$ coordinates of points in $V(I)$ whose $T$ coordinate is $\alpha$. So $T$ separates $V(I)$ if and only if for every $\alpha$ root of $f_k$, 
        \begin{equation*}
            g_k(\alpha,X) = a_{k,k}(\alpha)(X-\beta)^k.
        \end{equation*}

        As we've seen in Proposition~\ref{propositionmaintheorembivariate}, $\beta=-\frac{a_{k,k-1}(\alpha)}{ka_{k,k}(\alpha)}$ from comparing the coefficients of degree $k-1$. But we can compare all coefficients, and in order to obtain a simpler equation we look at the ratio of successive coefficients, to get that $g_k(\alpha,X)$ has exactly one root if and only if
        \begin{equation*}
            \frac{a_{k,i+1}(\alpha)}{a_{k,i}(\alpha)} = \frac{a_{k,k}(\alpha)\left(\begin{array}{c}k\\i+1\end{array}\right)(-\beta)^{k-i-1}}{a_{k,k}(\alpha)\left(\begin{array}{c}k\\i\end{array}\right)(-\beta)^{k-i}} = \frac{k-i}{i+1}\cdot\frac{1}{-\beta}=\frac{k-i}{i+1}\frac{ka_{k,k}(\alpha)}{a_{k,k-1}(\alpha)}
        \end{equation*}

        which, to put in a single line, is the same as
        \begin{equation*}
            a_{k,i+1}(\alpha)a_{k,k-1}(\alpha) = \frac{k-i}{i+1}ka_{k,k}(\alpha)a_{k,i}(\alpha).
        \end{equation*}

        Equality of these two expressions for all roots $\alpha$ of $f_k$ is equivalent by the Chinese Remainder Theorem to the following congruence:
        \begin{equation*}
            \frac{k-i}{i+1}ka_{k,k}a_{k,i}\equiv a_{k,i+1}a_{k,k-1}\mod f_k,
        \end{equation*}

        which is therefore true for all $k=1,\ldots,m$ and $i=0,\ldots,k-1$ if and only if $T$ separates $V(I)$.
    \end{proof}

Naturally these results lead first to the following algorithm in two variables.

    \begin{algorithm}\label{algorithmseparationtestbivariate}\caption{BivariateSeparationTest}
        \begin{algorithmic}[1]
            \Require A lexicographic Gr\"obner basis
            \State $G = \{f(T),g_1(T,X),\ldots,g_{m}(T,X)\}$ of $J\subseteq\KK[T,X]$ with
            \State $g_k=\sum\limits_{i=0}^{\partial_k}a_{k,i}(T)X^i$ (possibly with some $g_k=0$).
            \Ensure {\it True} if $T$ separates the zeroes of $J$, {\it False} otherwise. 
            \State $h_0\deff\overline{f}$
            \For{$k=1,\ldots,m$}
                \State $h_k\deff\gcd(h_{k-1},a_{k,\partial_k}),f_k\deff\frac{h_{k-1}}{h_k}$
                \For{$i=0,\ldots,\partial_k-1$}
                    \If{$\left(\begin{array}{c}\frac{\partial_k(\partial_k-i)}{i+1}a_{k,\partial_k}a_{k,i}\\ \neq \\ a_{k,\partial_k-1}a_{k,i+1}\,\operatorname{mod}f_{k}\end{array}\right)$}\label{separabilitycondition}
                        \State \Return {\it False}
                    \EndIf
                \EndFor
            \EndFor
            \State\Return {\it True}
        \end{algorithmic}
    \end{algorithm}

Algorithm \ref{LasVegasRadicalRURalgo} is a slight modification of Algorithm \ref{RadicalRURCandidatealgo}, which computes the RUR-candidate but incorporating the bivariate separation test: it represents the core of the Las Vegas strategy, solving Problem {\bf S}. It is based on the fact that if $T$ separates $V(I+\langle T-t\rangle\cap \mathbb{K}[T,X_i])$ for every $i=1..n$, then $T$ separates $V(I+\langle T-t\rangle)$ and thus $t$ separates $V(I)$.

    \begin{algorithm}\caption{LasVegasRadicalRUR}\label{LasVegasRadicalRURalgo}
        \begin{algorithmic}[1]
            \Require $M_{X_1},\ldots M_{X_n}$ the matrices of multiplication by each variable in $\frac{\mathbb{K} [X_1,\ldots,X_n]}{I}$ in some base $\mathcal{B}_0$ and a linear form $t=\sum_{i=1}^nt_iX_i\in\mathbb{K} [X_1,\ldots,X_n]$.
            \Ensure $\overline{f}(T)=0,X_i=\frac{f_i(T)}{\overline{f}'(T)}$, the reduced RUR of $V(\sqrt{I})$ associated to $t$ if $t$ separates $V(I)$, Fail otherwise
            \State Compute $\tilde{f},data:=\operatorname{MinimalPolynomial}(\sum_{i=1}^nt_iM_{X_i})$
            \State Compute $\overline{f}(T)$, the squarefree part of $\tilde{f}(T)$.
            \For{$i=1\ldots n$} 
                \State $G_i:={f}\cup \tmop{Coordinate}(M_{X_i},data)$
                \State {\bf If} not(BivariateSeparationTest($G_i$)) {\bf then} return(Fail,"$X_i$ not separated")
                \State $h_{i,1}X_i+h_{i,0}:=\tmop{BivariateParametrizationCandidate}(G_i)$  
                \State $f_{X_i}:=\frac{-1}{\deg(\overline{f})}h_{j,0}(T)h_{j,1}(T)^{-1}\overline{f}'\mod\overline{f}$
            \EndFor
        \end{algorithmic}
    \end{algorithm}

        It is easy to see that Algorithm \ref{algorithmseparationtestbivariate} runs $m\leq D-\delta+1$ times the main loop computing $\delta_k\leq \delta=\deg(f)$ elementary operations with polynomials of degree at most $\delta$ thus costing $O(\delta^2(D-\delta+1))$ operations in $\KK$. As this algorithm is call $n$ times in Algorithm \ref{LasVegasRadicalRURalgo}, we get:
     \begin{corollary}[Complexity of Algorithm \ref{LasVegasRadicalRURalgo}]
         Algorithm \ref{LasVegasRadicalRURalgo} Computes a RUR-Candidate of $V(I)$ associated to a given linear form $t$ or returns Fail is $t$ does not separate $V(I)$ in $\tilde{O}(D^2(D + n(D-\delta+1)))$ operations in $\KK$.
    \end{corollary}



    \subsection{Influences of separating elements}
    
    This is where the Las Vegas aspect of our algorithm materializes: we have no efficient deterministic way of choosing a {\it separating} linear form, even though one chosen at random will likely be separating. 
    
    For systems with rational coefficients, the choice of the separating linear form $t$ will influence the sparsity of the multiplication matrix $M_t$ as well as the size of the result and therefore the computation time. This choice also influences the difficulty of using the result, for example, the numerical approximation of the system's solutions from the reduced RUR associated with $t$.

 These various influences cannot be observed on the theoretical complexity bounds and therefore represent one of the main goals of this experimental section.

\subsubsection{Influence on the computation time and size of the result}

We have implemented Algorithm~\ref{LasVegasRadicalRURalgo} (and all necessary sub-algorithms) in Julia language~\cite{julia} in the {\tt RationalUnivariateRepresentation.jl} package. The source code of our implementations is freely available at
\begin{center}
    \url{https://gitlab.inria.fr/newrur/code/}
\end{center}

In this code, we have implemented a basic version of the FGLM algorithm using a straightforward handmade set of linear algebra functions. We used the package {\tt Groebner.jl}~\cite{GroebnerJL} for computation of Gröbner bases for the degree reverse lexicographic ordering.
We used the package {\tt Nemo.jl}~\cite{FHHJ17} for univariate polynomial arithmetic. 

For systems with rational coefficients, the implementation uses multi-modular tracing to speed up the computation of the modular images of the Gr{\"o}bner bases~\cite{T89}. The implementation first performs a search for a separating linear form modulo a prime number using Algorithm~\ref{LasVegasRadicalRURalgo}, fixes the linear form, and uses Algorithm \ref{RadicalRURCandidatealgo} for computing parametrizations over integers modulo a prime for multiple different prime numbers. As we currently do not use any theoretical bound on the size of the result, our implementation for systems with rational coefficients is Monte Carlo; it uses rational number reconstruction and halts when reconstruction succeeds~\cite{MCA}. On the other hand, the implementation over integers modulo a prime is Las Vegas and closely follows the algorithms in Sections~\ref{section:2},~\ref{section:3}.


With our benchmarks, we study the effect of the choice of the linear form on the RUR and its computation. Running modulo a prime number allows us to measure the effect of the choice of the linear form on the sparsity of the matrices used (and thus the efficiency of some parts of the computation), while running the multi-modular algorithm for systems with rational coefficients measure the effect of the choice of the linear form on the bitsize of the result.

Our benchmark suite consists of zero-dimensional systems over the rationals. The suite includes systems with cyclic quotient algebra where no single indeterminate is separating and the general case of non-radical systems. 
The sources of benchmark systems are available at \url{https://gitlab.inria.fr/newrur/code/Data/Systems}.

\begin{remark}[On other implementations]
As we will see in this section, computation times and the use of a RUR depend on many parameters, particularly the choice strategies for separating elements, making detailed analysis of computation times difficult.

Although we prefer not to compare our implementation too closely to other software, to provide context, we emphasize that it is competitive, despite its compactness (1,500 lines of code). For example, we observed that, on most of the examples tested in this section, our implementation is competitive with state-of-the-art implementations such as {\tt msolve}~\cite{msolve}.
\end{remark}

In the sequel, all computations are run using the packages out of the box on a single thread. Our setup is:
\begin{itemize}
    \item Julia 1.11.3 with \verb+Nemo.jl v0.48.4+, \verb+Groebner.jl v0.9.0+
    \item PC (Linux) : \verb+RAM : 128 GB+ \verb+CPU: 32 × 13th Gen i9-13900+
\end{itemize}

\begin{table}[!htbp]
    \setlength{\tabcolsep}{4pt}
  \caption{Benchmark results for computing rational univariate representation over the rationals using a multi-modular algorithm. The running time includes the search for a separating element. The bitsize is computed as the largest combined size (i.e., $\operatorname{log}_2$) of the numerator and the denominator of a coefficient in the result. In the table, Dim. stands for the dimension of the quotient algebra, Fail stands for an incorrect result.}
  \centering
    \begin{tabular}{c|cc|rr|rr|rr}
    \hline
    \multirow{2}{*}{System} &
    \multicolumn{2}{c|}{System info} &
    \multicolumn{2}{c|}{Random $-10..10$} &
    \multicolumn{2}{c|}{Random $-100..100$} &
    \multicolumn{2}{c}{Certified}\\
    & Dim. & Type & Bitsize & Runtime & Bitsize & Runtime & Bitsize & Runtime
    \\
    \hline
    Reimer 6 & 576 & Cyclic & 3,940 & 3 s & 5,638 & 4 s & 1,924 & 2 s \\
    Reimer 7 & 2,880 & Cyclic & Fail & Fail & 26,146 & 3,498 s & 12,226 & 749 s \\
    Noon 6      & 717   & Cyclic    & Fail & Fail & 6,034   & 10 s & 4,087 & 10 s\\
    Noon 7      & 2,173   & Cyclic    & Fail & Fail & 18,805   & 541 s & 14,562 & 519 s\\
    SEIR 36 & 14 & Cyclic & 221,191 & 21 s & 221,272 & 21 s & 136,990 & 10 s \\
    Crauste 2   & 128   & Cyclic    & 774,747 & 474 s & 774,967 & 863 s & 317,536 & 148 s\\
    Schwarz 11 & 2,048 & Cyclic & 16,013 & 357 s & 22,403 & 580 s & 12,010 & 232 s\\
    Goodwin & 40 & Cyclic & 69,200 & 5,606 s & 69,532 & 4,559 s & 19,652 & 1 s \\
    Fab 4       & 1,296 & General   & 29,736 & 174 s & 32,837 & 175 s & 28,056 & 178 s \\
    Ro. 5 sq      & 3,840 & General & Fail & Fail & 644 &  104 s  & 193 & 58 s \\
    Ch. 6 sq & 2,048 & General & 6,503 & 189 s & 6,556 & 177 s & 3,577 & 69 s \\
    No. 5 sq      & 7,456   & General & Fail & Fail & 1,610 & 7,376 s & 1,107  & 6,160 s \\
    Re. 5 sq  & 4,608 & General & 599 & 692 s & 1,093   & 1,379 s   & 363  & 763 s \\
    Ka. 6 sq & 8,192 & General & 631 & 2,156 s & 768 & 2,539 s & 382 & 1,008 s \\
    \hline
    \end{tabular}
  \label{table:benchmarks}
\end{table}

\begin{remark}[On Goodwin example]
In the Goodwin system, the difference in timings between a random linear form and a certified linear form is more than thousandfold. This can be explained by the fact that the Gr{\"o}bner basis of the original system contains several linear polynomials, hence, the computation of the basis of $I + \langle T - t \rangle$ for a dense $t$ requires auto-reduction and is not trivial. On the other hand, the certified separating form in that example is sparse.
\end{remark}

\begin{remark}[On reported bitsize]
In the table, an alternative way to measure the bitsize of the parametrization is to report the size of the coefficients of the parametrization after multipliying each polynomial by the least common multiple of its denominators so that coefficients become integers. In several (but not all) examples, this would have halved the reported size. We chose to report the bitsize as the size of rational numbers because we feel it better represents the size of the output, which is important for the efficiency of rational number reconstruction.
\end{remark}

We consider two strategies for the search for a separating element:
\begin{itemize}
    \item Random choice. The candidate $t = t_1 X_1 + \ldots + t_n X_n$ is chosen by picking each $t_1,\ldots,t_n$ as random integers in the range from $-B$ to $B$ (except zero) for a fixed $B\in\mathbb{N}$.
    
    \item Certified choice (based on Algorithm~\ref{LasVegasRadicalRURalgo}). The candidate $t = t_1 X_1 + \ldots + t_n X_n$ is formed for several different choices of $t_1,\ldots,t_n$ and for each choice Algorithm~\ref{LasVegasRadicalRURalgo} is run to check if the candidate is a separating linear form. The search halts when Algorithm~\ref{LasVegasRadicalRURalgo} succeeds.
    
    In our implementation, we use the following installation of this strategy. Let $I \in \mathbb{K}[X_1,\ldots,X_n]$ be a zero-dimensional ideal. Let $t\deff X_{n-1} - X_n$. If $t$ is separating for $V(I)$ (as verified by Algorithm~\ref{LasVegasRadicalRURalgo}), then halt. Otherwise, let $X_i$ be the first (from the end) indeterminate such that $t$ is not separating for $V(I\cap\mathbb{K}[X_{i},\ldots,X_n])$. Set $t_i \deff t_i + 1$ and $t \deff t_{i}X_{i}+t$ and repeat the procedure with the new $t$.
\end{itemize}
In the first experiment, for each system we run the search for a separating form and then run the computation of a rational univariate representation using a multi-modular algorithm. Results are summarized in Table~\ref{table:benchmarks}. In the table, we include three strategies for the search for a separating form: random choice with absolute values of coefficients of linear form bounded by $10$, random choice with absolute values of coefficients of linear form bounded by $100$, and the certified strategy where the checks for a separating form are performed modulo a prime. For each system in the benchmark, we report:
\begin{itemize}
    \item Information about the system: the dimension of the quotient algebra, the type of the system;
    \item For each search strategy: the bitsize of the result computed as the largest combined size (i.e., $\operatorname{log}_2$ of the absolute value) of the numerator and denominator of a coefficient;
    \item For each search strategy: elapsed running time, which includes the time for the search for a separating form (which contributes significantly only for the certified strategy);
\end{itemize}

From Table~\ref{table:benchmarks}, we make several observations.
The first thing to note is that in all examples the certified strategy finds a separating form that produces a parametrization with bitsize smaller compared to random linear forms, by at most 3 times in some examples. When picking the coefficients of a linear form randomly from $-10..10$, the coefficients of a parametrization are slightly smaller than when picking coefficients from $-100..100$. However, in several examples, the former produces incorrect result, meaning that the random linear form was not separating. As we shall see, the feature of the certified strategy is that it may produce relatively sparse separating forms, which seems to lead to smaller coefficients of a parametrization.

Note that the running time of the computation with the certified strategy is in some cases larger than the running time with a random pick. For example, for the Re. 5 sq example, the search with the certified strategy took several attempts to produce a separating form, and the size of the final parametrization was not large, which made certified strategy slower compared to the random choice. Generally, however, certified search produces parametrizations with smaller coefficients, which decreases the running time for some examples by up to 6 times.



\begin{table}[!htbp]
\setlength{\tabcolsep}{4pt}
  \caption{The sparsity of the multiplication matrix $M_t$ and the sparsity of the associated separating form $t = t_1 X_1 + \ldots + t_n X_n$ calculated as the number of nonzeros among $t_1,\ldots,t_n$ for two strategies of picking separating form: random choice and certified choice.}
  \centering
    \begin{tabular}{c|rr|rr}
    \hline
  \multirow{2}{*}{System} & \multicolumn{2}{c|}{Random} & \multicolumn{2}{c}{Certified}\\
  & { Sparsity of $M_T$} & { Sparsity of $t$} & { Sparsity of $M_T$} & { Sparsity of $t$}\\
    \hline
    Reimer 6    & $0.43$    & 6/6 & $0.43$ & 4/6  \\
    Reimer 7    & $0.39$    & 7/7 & $0.39$ & 5/7  \\
    Noon 6      & $0.10$    & 6/6 & $0.10$ & 5/6  \\
    Noon 7      & $0.06$    & 7/7 & $0.06$ & 6/7  \\
    SEIR 36     & $1.00$    & 68/68 & $0.87$ & 1/68 \\
    Crauste 2   & $0.48$    & 43/43 & $0.36$ & 3/43  \\
    Schwarz 11  & $0.28$    & 11/11 & $0.06$ & 3/11 \\
    Goodwin     & $0.10$    & 22/22 & $0.08$ & 4/22 \\
    Fab 4       & $0.24$    & 4/4 & $0.24$ & 4/4  \\
    Ro. 5 sq     & $0.07$    & 5/5 & $0.07$ & 4/5  \\
    Ch. 6 sq     & $0.38$    & 6/6 & $0.22$ & 1/6   \\
    No. 5 sq     & $0.17$    & 5/5 & $0.16$ & 4/5   \\
    Re. 5 sq     & $0.31$    & 6/6 & $0.30$ & 3/6  \\
    Ka. 6 sq     & $0.30$    & 6/6 & $0.13$ & 1/6   \\
        \hline
    \end{tabular}
  \label{table:sparsity}
\end{table}
    
In the next experiment, for each system, we calculate the sparsity of the matrix $M_t$ of multiplication by $t$ in the quotient algebra and the sparsity of $t$, that is, the number of nonzeros among $t_1,\ldots,t_n$, for a fixed separating form $t = t_1 X_1 + \ldots + t_n X_n$. We conduct this experiment for two strategies: a random choice of separating form and a certified choice of separating form. 

We report results in Table~\ref{table:sparsity}. For each system in the benchmark, the table contains:
\begin{itemize}
    \item For each search strategy: the sparsity of $M_t$, computed as the number of nonzeros divided by the number of entries.
    \item For each search strategy: the sparisty of $t$.
\end{itemize}
We observe that the certified strategy tends to produce sparse separating elements in many examples. As we have seen in the first experiment, this seems to lead to parametrizations with smaller coefficients. At the same time, the sparsity of the matrix $M_t$ generally does not decrease dramatically for sparse separating forms.


\begin{remark}[On sparsity and other search strategies]
One natural question seems to be: for a given system, what is the most sparse separating form. A partial answer can be obtained by enumerating all supports in $2^{\{1,\ldots,n\}}$, generating a random linear form with a fixed support, and checking if it is separating using Algorithm~\ref{LasVegasRadicalRURalgo}.
This procedure yields a separating form that is more sparse than the one obtained with the certified strategy for the following systems: Crauste 2, Goodwin, Schwarz 11.
However, we do not use this procedure as the default in the implementation, because, as is, the costly search process outweighs the benefits from sparsity.
\end{remark}

\subsubsection{Influence on real root isolation}

We now study the effect that the size of coefficients in the parametrization has on the efficiency of real root isolation, which is a typical step in the solution of polynomial systems via RUR.

\begin{table}[!htbp]
\setlength{\tabcolsep}{4pt}
  \caption{The running time of real root isolation for two choices of separating form: random and certified.}
  \centering
    \begin{tabular}{crr}
    \hline
    System 
    & \begin{tabular}{c}
         Runtime\\(Random)  
    \end{tabular} 
    & \begin{tabular}{c}
         Runtime\\(Certified)  
    \end{tabular} \\
    \hline
    Reimer 6 & 5 s & 3 s \\
    Reimer 7 & 3,262 s & 2,878 s \\
    Noon 6 & 1 s & 1 s \\
    Noon 7 & 7 s & 7 s \\
    Crauste 2 & 10 s & 4 s \\
    Fab 4 & 2 s & 2 s \\
        \hline
    \end{tabular}
  \label{table:isolation}
\end{table}

In Table~\ref{table:isolation}, for some of the systems considered previously, we report the total running time for isolating the roots of the minimal polynomial and for finding the solutions for all coordinates at required precision by backward substitution. We use the root isolation routine exported by the Julia package PACE.jl, which implements algorithms from~\cite{10.1145/2930889.2930937}. We request 50 bits of precision in all instances. Some of the systems previously considered had short runtimes for root isolation (i.e., less than a second) and are not reported in Table~\ref{table:isolation}. We observe that the size of the RUR coefficients does not influence too much the runtime of root isolation, which is likely because the most important information for this is the minimal distance between roots of the first RUR polynomial as well as the flatness of the rational functions in order to accurately evaluate the coordinates without much precision.


    \section{Conclusion}

    \paragraph{}In this paper, we pointed out previously unfound relations between Gr\"obner bases and rational parametrizations of zero-dimensional systems. From these results, we proposed a general Las Vegas algorithm for solving such systems over arbitrary fields by computing a separating linear form and the associated reduced Rational Univariate Representation. Our algorithm has a better complexity than previous Las Vegas solutions in the general case and is still competitive in specific cases like ideals in shape position.
    
    For the case of systems with rational coefficients, we implemented a multi-modular strategy making use of our main function for computing RURs in $\frac{\mathbb{Z}}{p\mathbb{Z}}$ with $p$ prime. As we did not use any bound on the size of the result, the reconstruction of the rational coefficients is probabilistic and thus the implementation is Monte-Carlo. We verified that our implementation is competitive with state of the art algorithms (which are all Monte-Carlo), and used our implementation to measure the effect of the choice of the separating form on the computation time as well as on the size of the result. 
    
     Since almost all objects computed (several of the Gr\"obner bases and the entire parametrization) are dependent on the linear form one chooses, it would be reasonable to expect that this choice can give radically different results. This was only partially verified experimentally: in several examples our certified strategy was faster to find a separating linear form and compute the associated RUR, but often not by much, as seen in Table~\ref{table:benchmarks}. What was unexpected was the apparent lack of influence the sparsity of the chosen linear form on the size of the RUR or the sparsity of the multiplication matrices in most cases. The exception were the systems SEIR 36, Schwarz 11, Ch. 6 sq, Ka. 6 sq, where a significantly sparse linear form yields a noticeably sparser multiplication matrix (a ratio varying from $1.15$ to $4.67$). 
    
    We have observed that the use of RURs, for example for the isolation of real roots of systems, is not influenced significantly by the choice of separating elements. The reason might be twofold: on one hand the size of the coefficients does not have much influence since the isolation algorithm essentially uses floating-point arithmetic, on the other hand it is rather the distance between the roots and the maximum modulus of the roots which significantly vary the calculation times and our choice of separating elements does not take this into account. On the other hand, the flatness of the coordinates functions seems to play a role in the precision required for evaluating accurately the coordinate functions.
     
     Choosing better separating elements for optimizing the computation time, size and use of the RUR is thus a challenging open problem we would like to tackle in future contributions. 
     
     In this article we first decompose the system, but we reconstruct a unique RUR in order to compute \textit{the} reduced RUR associated to a given separating form. Modifying slightly the separation test, we can easily decompose the system in several subsystems, some which are well separated by the current linear form, others with a default of separation for at least one coordinate so that at the next step we can concentrate only on the non well separated subsystems with some extraneous information about the non separated coordinates. Instead of getting a unique reduced RUR associated to a unique separating form we will get a decomposition into RURs associated to different separating forms. A challenge will then be to show that the complexity of such an algorithm is at least as good as the one of the present article.

    \tolerance=9000
    \bibliographystyle{alpha}
    \bibliography{bibliography.bib}

\appendix

\section{Complement for reviewers}

As stated in the article, we do not wish to publish a comparison with existing competing implementations. There are several reasons for this. As we showed in the experimental section, a change in the separating linear form can have uncontrolled effects on computation time, output size, and object utilization. In fact, it can even significantly impact the computation of Gröbner bases, which is not the subject of this contribution. Thus, time comparisons are difficult to analyze between the tools (Gröbner bases), the environment (programming languages and compilers), the algorithms, and arbitrary choices whose consequences are uncontrollable (separating elements).

Nevertheless, it is legitimate for a reviewer to ask to place our implementation among existing implementations. We indicated in the main text that our implementation was competitive, and in this appendix we provide benchmark results to support this claim. We compare with a solution that self-evaluates as one of the best existing (\verb+msolve+ - \cite{msolve}) on 2 different hardware architectures (we compared on two architectures because \verb+msolve+ contains assembler code which could add an additional parameter in analysis of timings). We used \verb+msolve+ version 0.7.4.

All the experiments can be reproduced.

For msolve, we use directly the C code to be sure to run the best possible variant, for example the one using assembly AVX512 instructions.

Our code as well as the input systems are available at \verb+https://gitlab.inria.fr/newrur/code+. There is a rather efficient and very simple variant written in Maple and a more efficient variant written in Julia. None are using architecture dependent tricks.

In the following tables:

\begin{itemize}
    \item timings are expressed in seconds;
    \item \verb+Msolve-Param-C code+ is the computation of a parameterization with msolve using the C code (not the Julia interface);
    \item \verb+PACE-Param-Julia+ is the computation of a reduced RUR of the radical with our algorithm written in Julia;
    \item \verb+Msolve-Roots-C code+ is the computation of the real roots of the system including the computation of a parameterization with msolve using the C code (not the Julia interface);
    \item \verb+PACE-Roots-Julia+ is the computation of the real roots of the system using RS library written in C after the computation of a reduced RUR of the radical  with our algorithm written in Julia
    \item \verb+TYPE+ indicates whether each system is in shape position (S), has a cyclic quotient algebra (C) or none of those (G).
\end{itemize}

\includegraphics[width=\columnwidth]{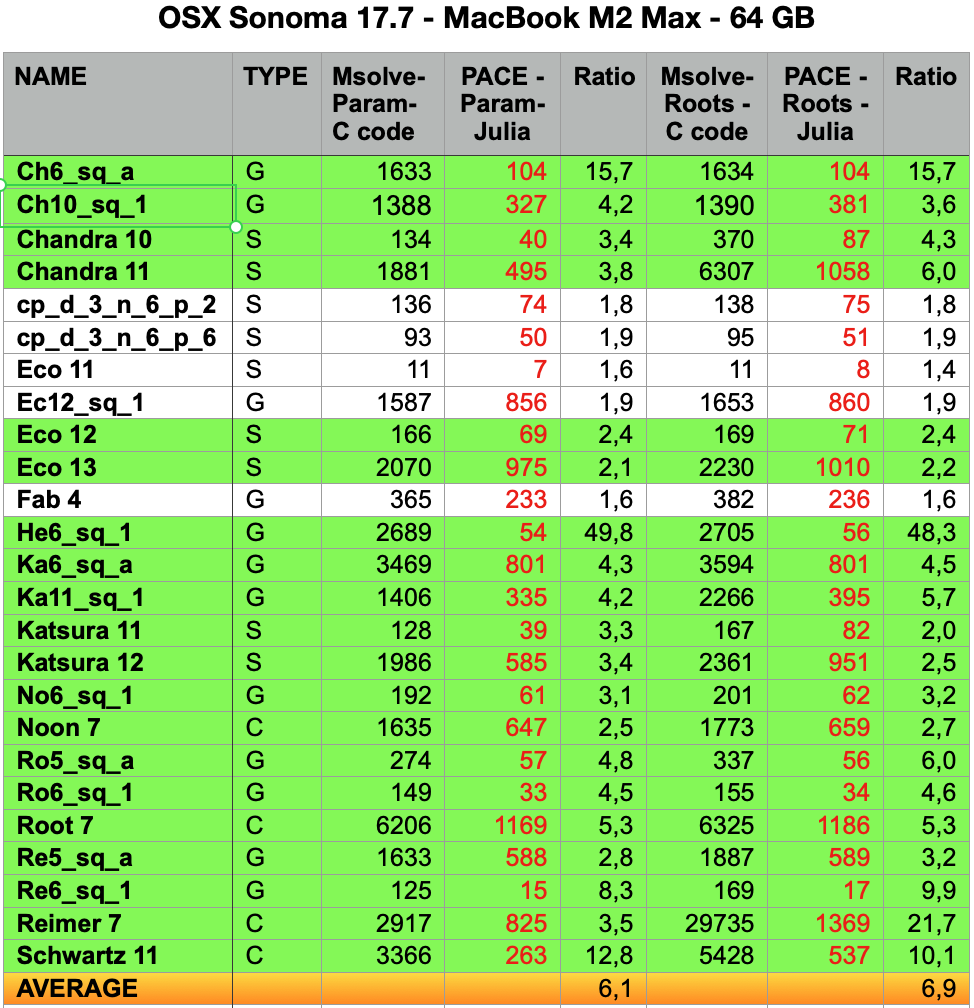}

\includegraphics[width=\columnwidth]{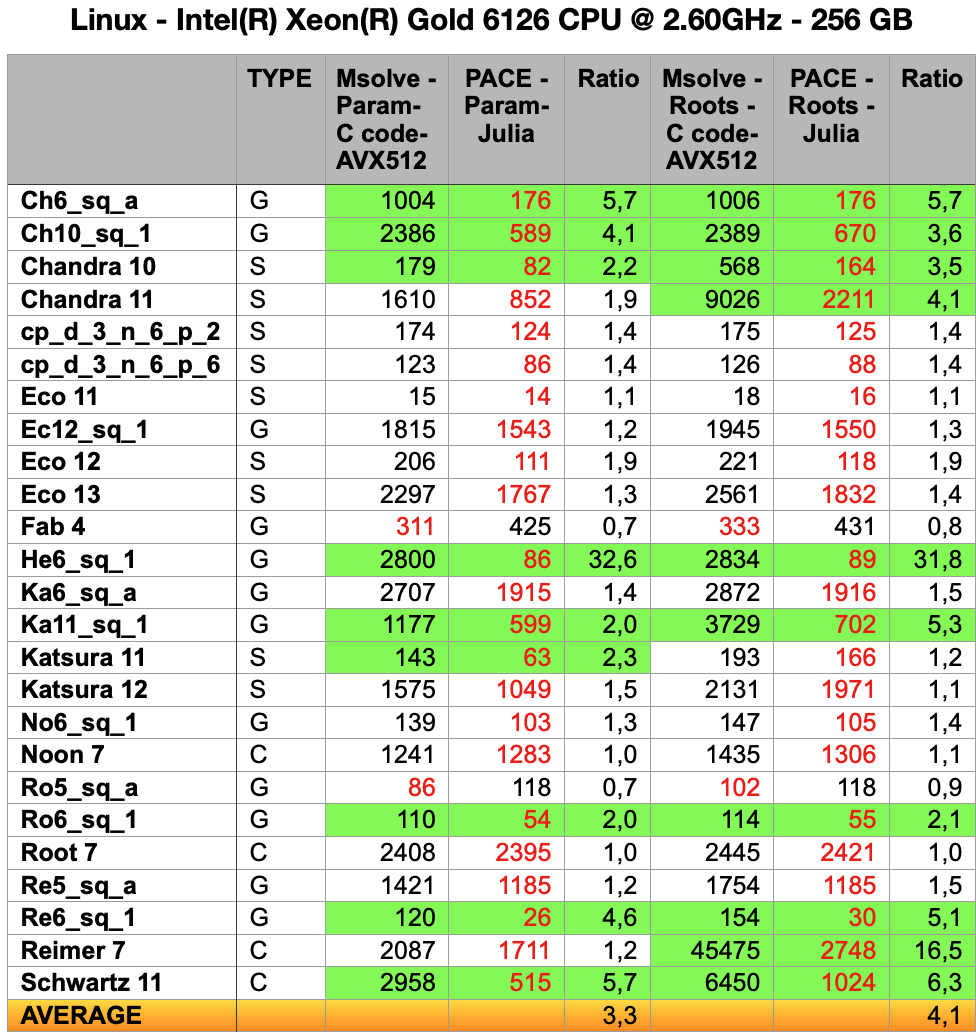}

\end{document}